\documentstyle[psfig]{elsart}

\newcommand{\lya}{{Ly $\alpha$}}
\newcommand{\gray}{{$\gamma$-ray}}
\newcommand{\grays}{{$\gamma$-rays}}
\newcommand{\ie}{{\it i.e.}}

\newcommand{\apj}{{Astrophys. J.}}
\newcommand{\apjl}{{Astrophys. J. (Lett.)}}
\newcommand{\apjs}{{Astrophys. J. Supp.}}

\newcommand{\xxiicrc}{{21st Internat. Cosmic Ray Conf.}}
\newcommand{\xxvicrc}{{25th Internat. Cosmic Ray Conf.}}
\newcommand{\aap}{{Astron. and Astr.}}
\newcommand{\mic}{$\mu$m}
\newcommand{\aaps}{{Astron. and Astr. Suppl.}}

\begin{document}

\begin{frontmatter}

\title{Intergalactic Extinction of High Energy Gamma-Rays}

\thanks[talk]{Invited Talk Presented at the Workshop on TeV Astrophysics 
of Extragalactic Sources, Cambridge, MA, October 1998; to be published in
the Proceedings.}

\author{F.W. Stecker}

\address{Laboratory for High Energy Astrophysics, NASA Goddard Space Flight 
Center, Greenbelt, MD 20771, USA}

\begin{abstract}
We discuss the determination of the intergalactic pair-production absorption 
coefficient as derived by
Stecker and De Jager by making use 
of a new empirically based calculation of the spectral energy distribution of 
the intergalactic infrared radiation field as given by Malkan and Stecker.
We show that the results of the Malkan and Stecker calculation agree well with
recent data on the infrared background.
We then show that previous spectral data from observations
of Mrk 421 and Mrk 501 are consistent with the
amount of intergalactic absorption predicted by Stecker and De Jager and
that the new HEGRA observations of the flaring spectrum of Mrk 501 presented at
this conference actually appear to
show the amount of intergalactic absorption which we predict. 
As a further test for intergalactic absorption, we give a predicted spectrum, 
with absorption included, 
for PKS 2155-304. This XBL lies at a redshift of 0.12, the highest 
redshift source yet observed at an energy above 0.3 TeV.
This source should have its spectrum steepened by $\sim 1$ in its spectral
index between $\sim 0.3$ and $\sim 3$ TeV and should show an absorption 
cutoff above $\sim 6$ TeV.
We also discuss the determination 
of the \gray\ opacity at higher redshifts (out to $z=3$), following the 
treatment of 
Salamon and Stecker.

\end{abstract}


\end{frontmatter}

\section{Introduction}

Very high energy \gray\ beams from blazars can be used to 
measure the intergalactic infrared radiation field, since 
pair-production interactions of \grays\ with intergalactic IR photons 
will attenuate the high-energy ends of blazar spectra \cite{sds92}. 
In recent years, this concept has been used successfully to place upper limits 
on the the intergalactic IR field (IIRF) \cite{sd93}
- \cite{bil98}.
Determining the (IIRF), in turn, allows us to 
model the evolution of the galaxies which produce it. 
As energy thresholds are lowered 
in both existing and planned ground-based
air Cherenkov light detectors \cite{knp}, cutoffs in the \gray\ spectra of 
more distant blazars are expected, owing to extinction by the IIRF. These
can be used to explore the redshift dependence of the 
IIRF \cite{ss97}, \cite{ss98}. 

There are now 66 ``grazars'' ($\gamma$-ray blazars) which have been 
detected by the {\it EGRET} team
\cite{3egret}. These sources, optically violent variable quasars
and BL Lac objects, have been detected out to a redshift greater that 2.
Of all of the blazars detected by {\it EGRET}, only the low-redshift 
BL Lac, Mrk 421 ($z = 0.031$), has been seen by
the Whipple telescope \cite{punch92}. The fact that the Whipple team did not 
detect the much brighter {\it EGRET} source, 3C279, at TeV energies
\cite{vac90}, \cite{ker93} is consistent with the predictions of a
cutoff for a source at its much higher redshift of 0.54 \cite{sds92}.
So too are the further detections of three other close BL Lacs 
($z < 0.12$), {\it viz.}, Mrk 501 ($z = 0.034$) \cite{quinn96}, 1ES2344+514 
($z = 0.044$)\cite{cat98}, and PKS 2155-304 ($z = 0.117$) \cite{chad98} which 
were too faint at GeV energies to be seen by {\it EGRET}\footnotemark\footnotetext{PKS 2155-304 was seen in one observing period by
{\it EGRET} as reported in the Third EGRET Catalogue \cite{3egret}}.

\section{The Opacity of Intergalactic Space Owing to the IIRF}

The formulae relevant to absorption calculations involving pair-production 
are given and discussed in Ref. \cite{sds92}.
For $\gamma$-rays in the TeV energy range, the pair-production cross section 
is maximized when the soft photon energy is in the infrared range:
\begin{equation} 
\lambda (E_{\gamma}) \simeq \lambda_{e}{E_{\gamma}\over{2m_{e}c^{2}}} =
2.4E_{\gamma,TeV} \; \; \mu m 
\end{equation}
where $\lambda_{e} = h/(m_{e}c)$ 
is the Compton wavelength of the electron.
For a 1 TeV $\gamma$-ray, this corresponds to a soft photon having a
wavelength  near the K-band (2.2\mic). (Pair-production interactions actually
take place with photons over a range of wavelengths around the optimal value as
determined by the energy dependence of the cross section; see eq. (6)).) 
If the emission spectrum of
an extragalactic source extends beyond 20 TeV, then the extragalactic
infrared field should cut off the {\it observed} spectrum between $\sim
20$ GeV and $\sim 20$ TeV, depending on the redshift of the source \cite{ss97},
\cite{ss98}.

\section{Absorption of Gamma-Rays at Low Redshifts}

Stecker and De Jager \cite{sd98} (hereafter SD98) have recalculated the 
absorption coefficient of intergalactic
space using a new, empirically based calculation
of the spectral energy distribution (SED) of intergalactic low energy 
photons by Malkan and Stecker \cite{ms98} (hereafter MS98) 
obtained by integrating luminosity dependent infrared spectra of galaxies
over their luminosity and redshift distributions.
After giving their results on the \gray\ optical depth as a function of energy 
and redshift out to a redshift of 0.3, SD98 applied their calculations by
comparing their results with the spectral data on Mrk 421 \cite{mch97} and 
spectral data on Mrk 501 \cite{ah98}. 

SD98 make the reasonable simplifying assumption 
that the IIRF is basically in
place at a redshifts $<$ 0.3, having been produced primarily at higher
redshifts \cite{ss97}, \cite{ss98}, \cite{mad96}. 
Therefore SD98 limited their calculations to $z<0.3$. (The calculation of
\gray\ opacity at higher redshifts \cite{ss97},\cite{ss98} will be discussed
in the next section.)

\begin{figure}
\centerline{\psfig{file=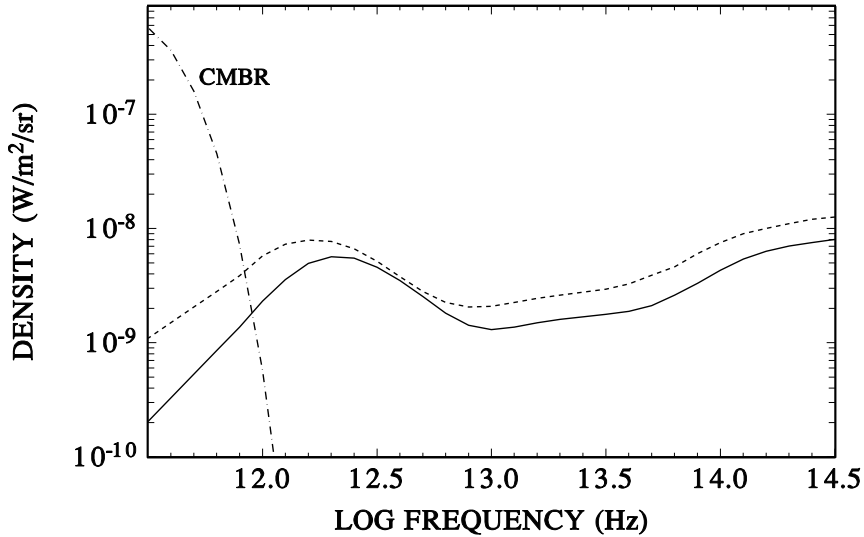,width=13.0truecm}}
\vspace{-8.0truecm}
\caption{The spectral energy distribution (SED) of the extragalactic 
IR radiation calculated
by Malkan and Stecker \protect\cite{ms98} with the 2.7 K cosmic background radiation 
spectrum added. The solid line (lower IIRF curve) and the dashed line 
(higher IIRF curve) correspond to the middle and upper curves calculated by 
Malkan and Stecker with redshift-evolution assumptions as 
described in the text.}
\label{IRSED}

\end{figure}

SD98 assumed for the IIRF, two of the SEDs given in MS98 \cite{ms98}
(shown in Figure \ref{IRSED}).
The lower curve in Figure \ref{IRSED} (adapted from MS98) assumes evolution 
out to $z=1$, whereas the upper curve assumes evolution out to $z=2$.
Evolution in stellar emissivity is expected to level off or decrease
at redshifts greater than $\sim 1.5$ \cite{mad96}-\cite{st98} 
so that the two curves in Fig. 1 may be considered to be lower and upper 
limits, bounding the expected IR flux.
Using these two SEDs for the IIRF, SD98
obtained parametric expressions for $\tau(E,z)$ for $z<0.3$, taking a 
Hubble constant of $H_o=65$ km s$^{-1}$Mpc$^{-1}$  
\cite{grat97}. 

\begin{figure}
\centerline{\psfig{file=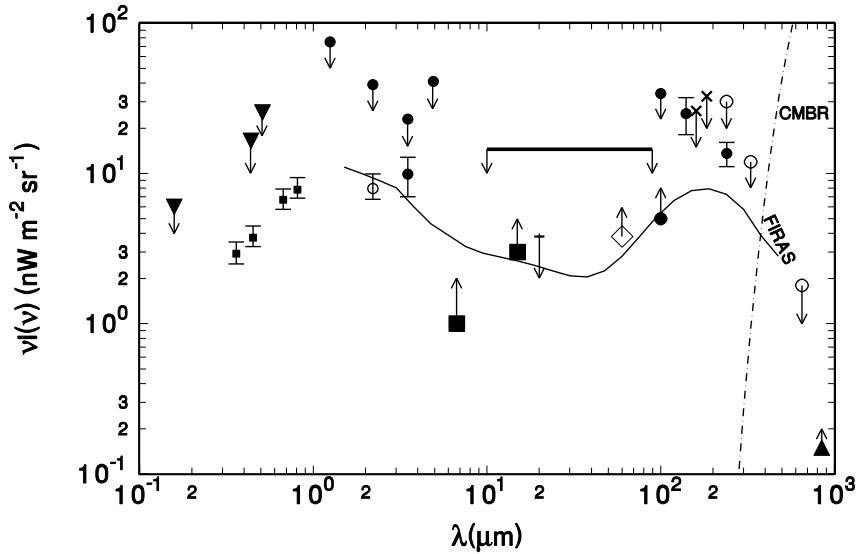,width=13.0truecm}}
\vspace{-8.0truecm}
\caption{The upper infrared SED from Malkan and Stecker compared with 
observational data and other constraints (courtesy O.C. De Jager).}
\label{irdata}
\end{figure}

The results of MS98 \cite{ms98} generally agree well with very recent 
{\it COBE} data.\footnotemark\footnotetext{The derived {\it COBE} point
at 140 \mic\ appears to be inconsistent with all 
calculated IIRF SEDs. It is also inconsistent with the spectrum of
Mrk 501 (Konopelko, these proceedings), since it would imply
a \gray\ optical depth $\sim 6$ at 20 TeV.}
and with lower limits from galaxy counts and other 
considerations \cite{pug96}  - \cite{dwe98b}. 
The results of MS are also in agreement with upper limits obtained from
TeV \gray\ studies \cite{sd93} - \cite{bil98}.
This agreement is illustrated in Figure \ref{irdata} which shows the 
upper SED curve from MS98 in comparison with various data and limits.

The double-peaked form of the SED of the IIRF requires
a 3rd order polynomial to approximate the opacity $\tau$ in a parametric form.
SD98 give the following approximation:
\begin{equation}
log_{10}[\tau(E_{\rm TeV},z)]\simeq\sum_{i=0}^3a_i(z)(\log_{10}E_{\rm TeV})^i
\;\;{\rm for}\;\;1.0<E_{\rm TeV}<50,
\end{equation}
where the z-dependent coefficients are given by
\begin{equation}
a_i(z)=\sum_{j=0}^2a_{ij}(\log_{10}{z})^{j}.
\end{equation}
Table 1 gives the numerical values for 
$a_{ij}$, with $i=0,1,2,3$, and $j=0,1,2$. The numbers before the
brackets are obtained using the lower IIRF SED shown in Figure 1; The
numbers in the brackets are obtained using the higher IIRF SED.
Equation (2) approximates $\tau(E,z)$ to within 10\% for all values of z and E 
considered.

\vspace{1em}

\begin{tabular}{|c|r|r|r|r|}
\multicolumn{5}{c}{\bf Table 1: Polynomial coefficients $a_{ij}$}\\
\hline
$j$ & $a_{0j}$ & $a_{1j}$ & $a_{2j}$ & $a_{3j}$ \\ \hline
0&1.11(1.46) &-0.26(~0.10) &1.17(0.42) &-0.24(~0.07)\\
1&1.15(1.46) &-1.24(-1.03) &2.28(1.66) &-0.88(-0.56)\\
2&0.00(0.15) &-0.41(-0.35) &0.78(0.58) &-0.31(-0.20)\\ \hline
\end{tabular}

\vspace{1em}

\begin{figure}[t]
\centerline{\psfig{file=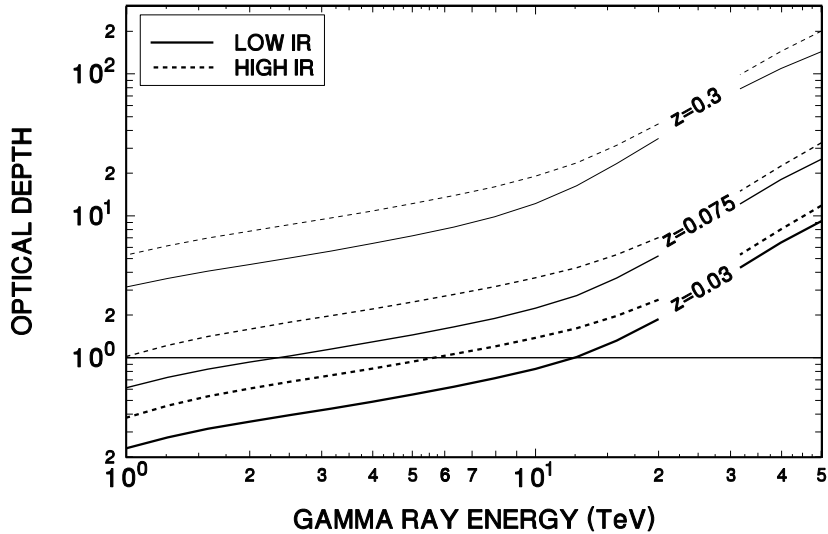,width=13.0truecm}}
\vspace{-8.0truecm}
\caption{
Optical depth versus energy for \grays\ originating at various redshifts
obtained using the SEDs corresponding to the lower IIRF (solid lines) and 
higher IIRF (dashed lines) levels shown in Fig. \protect\ref{IRSED} (from SD98).}
\label{lowztau}
\end{figure}

Figure \ref{lowztau} shows the results of the SD98 calculations 
of the optical depth for various energies and redshifts up to 0.3.

\begin{figure}[t]
\centerline{\psfig{file=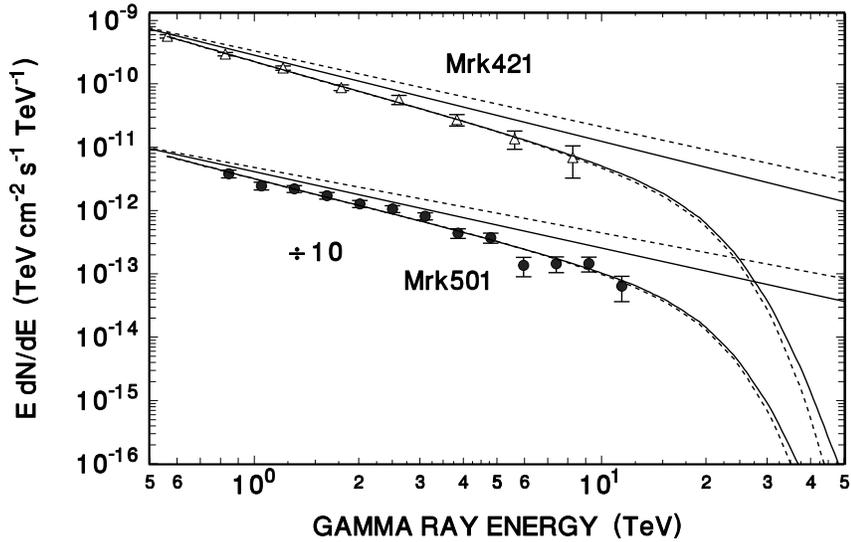,width=13.0truecm}}
\vspace{-8.0truecm}
\caption{
The observed spectra of Mrk 421 (open triangles) \protect\cite{mch97}
and Mrk 501 (solid circles - spectrum divided 
by 10) \protect\cite{ah98}. Best-fit absorbed spectra (of the form $KE^{-\Gamma}\exp(-\tau(E,z=0.03))$)
and implied unabsorbed spectra ($KE^{-\Gamma}$) for both sources
are shown for $\tau$ corresponding to
the lower IIRF SED (solid lines; $\Gamma=2.36$ and 2.2 for Mrk 421 and 
Mrk 501 respectively) and higher IIRF SED (dashed lines;
$\Gamma=2.2$ and 2.03 for Mrk 421 and Mrk 501 respectively) (from SD98).}
\label{mrkfig}
\end{figure}

Figure \ref{mrkfig} shows observed spectra 
for Mrk 421 \cite{mch97} and Mrk 501 \cite{ah98} 
in the flaring phase, compared with best-fit spectra of the form
$KE^{-\Gamma}\exp(-\tau(E,z=0.03))$, with $\tau(E,z)$ given by the two 
appropriate curves shown in Figure \ref{lowztau}. Because 
$\tau < 1$ for $E<10$, 
TeV, there is no obvious curvature in the
{\it differential} spectra below this energy; rather, we obtain a
slight steepening in the power-law spectra of the sources as a result 
of the weak absorption. This result implies that the 
{\it intrinsic} spectra of the sources should be harder by 
$\delta \Gamma \sim$ 0.25 in the 
lower IRRF case, and $\sim$ 0.45 in the higher IIRF case. 

\vskip 2.0truecm

The SD98 results for the absorption coefficient as 
a function of energy do not differ dramatically from those obtained
previously \cite{mp96}, \cite{sd97}; 
however, they are more reliable because they are
based on the empirically derived IIRF given by MS98, whereas all previous
calculations of TeV $\gamma$-ray absorption were based on theoretical modeling
of the IIRF. 
The MS98 calculation was based on data from nearly 3000
IRAS galaxies. These data included (1) the luminosity dependent infrared SEDs
of galaxies, (2) the 60$\mu$m luminosity function of galaxies and, (3) 
the redshift distribution of galaxies.
 
The advantage of using empirical
data to construct the SED of the IIRF, as done in MS98, is 
particularly indicated in the mid IR range. In this region of the spectrum,
galaxy observations indicate more flux from warm dust in galaxies than that
taken account of in more theoretically oriented models 
({\it e,g,} Primack, \etal, these
proceedings). As a consequence, the mid-IR ``valley'' between the cold dust 
peak in the far IR and cool star peak in the near IR is filled in more in the 
MS98 results and is
not as pronounced as in previously derived models of the IR background SED.
Such other derived SEDs are in conflict with recent lower
limits in the mid-IR derived from galaxy counts (see Figure \ref{irdata}).

The SD98 calculations predict 
that intergalactic absorption should only 
slightly steepen the spectra of Mrk 421 and Mrk 501 below $\sim$ 10 TeV,
which is consistent with the data already in the published literature
(see Figure \ref{mrkfig}). The SD98 calculations further predict that
intergalactic absorption should turn over the spectra of these sources 
at energies greater
than $\sim$20 TeV (see Figure \ref{mrkfig}). Observations of these 
objects at large zenith angles, 
which give large effective threshold energies, may thus demonstrate 
the effect of intergalactic absorption.

The observed spectrum of Mrk 501 in the flaring phase has been newly extended 
to an energy of 24 TeV by observations of the {\it HEGRA} group. (These new 
data are not shown in Figure \ref{mrkfig} but are given in the paper of 
Konopelko in these proceedings.) The new {\it HEGRA} data are well fitted
by a source spectrum power-law of spectral index $\sim 1.8$ steepened at
energies above a few TeV by intergalactic absorption with the optical depth
calculated by SD98 (Konopelko, private communication).

\begin{figure}[t]

\centerline{\psfig{file=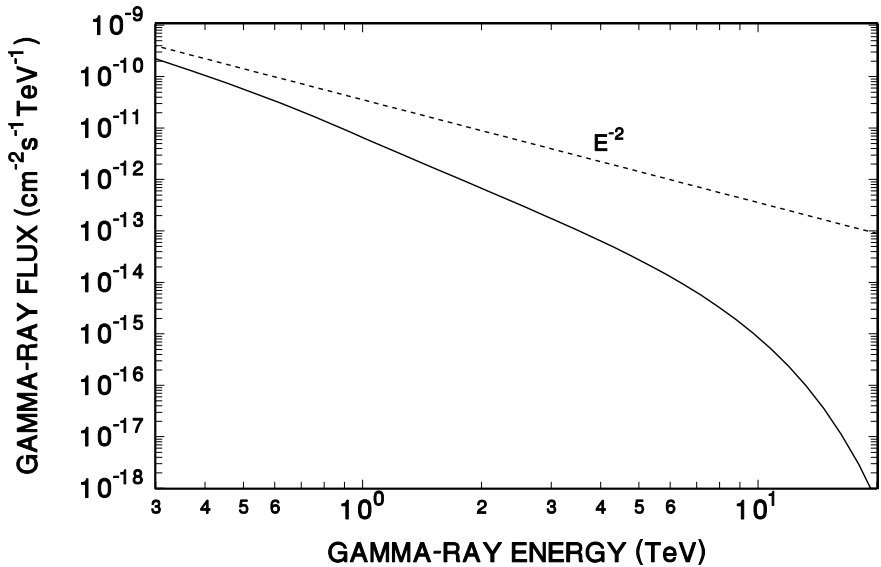,width=13.0truecm,angle=270}}
\vspace{2.0truecm}
\caption{Predicted differential absorbed spectrum, for PKS 2155-304 
(solid line) assuming an $E^{-2}$ differential source spectrum (dashed line) 
normalized
to the integral flux given in Ref. \protect\cite{chad98} (see text).}
\label{2155}
\end{figure}

Finally, we consider the source PKS 2155-304, an XBL located at a moderate
redshift of 0.117, which has been reported by the Durham group to have
a flux above 0.3 TeV of $\sim 4 \times 10^{-11}$ cm$^{-2}$ s$^{-1}$
\cite{chad98}, close to that predicted by a simple SSC model \cite{sds96}.
Using the SD98 absorption results for the higher IR SED in Figure \ref{IRSED}
and assuming an $E^{-2}$ source spectrum, we predict an absorbed (observed)
spectrum as shown in Figure \ref{2155}. As indicated in the figure, we find 
that this source should have its spectrum steepened by $\sim$ 1 in its spectral
index between $\sim 0.3$ and $\sim 3$ TeV and should show an absorption 
turnover above $\sim 6$ TeV. Observations of the spectrum of this source should
provide a further test for intergalactic absorption.

\section{Absorption of Gamma-Rays at High Redshifts}

We now discuss the absorption of 10 to
500 GeV \grays\ at high redshifts. 
In order to calculate such high-redshift absorption properly, it is
necessary to determine the spectral distribution of the intergalactic low 
energy photon background radiation as a function of redshift as realistically 
as possible out to frequncies beyond the Lyman limit. This calculation,
in turn, requires observationally based information on the evolution of the 
spectral energy distributions (SEDs) of IR through UV starlight from galaxies,
particularly at high redshifts. 

Conversely, observations of high-energy cutoffs in the
\gray\ spectra of blazars as a function of redshift, which may enable one to 
separate out intergalactic absorption from redshift-independent cutoff 
effects, could add to our knowledge of galaxy formation and early galaxy 
evolution. In this regard, it should be noted that the study of blazar spectra in the 10 to 300 GeV range is one of the primary goals of a next generation
space-based \gray\ telescope {\it GLAST (Gamma-ray Large Area Space Telescope)}
(Ref. \cite{bl96} and Gehrels, these proceedings) as well as {\it VERITAS} and 
other future ground based \gray\ telescopes. 

Salamon and Stecker \cite{ss98} (hereafter SS98) have calculated 
the \gray\ opacity as a function of both energy and redshift
for redshifts as high as 3 by taking account of the evolution of both the
SED and emissivity of galaxies with redshift (see section 4.2). 
In order to accomplish this, 
they adopted the recent analysis of Fall, \etal\  \cite{fall96} and 
also included the effects of metallicity evolution on galactic SEDs.
They then gave predicted \gray\ spectra 
for selected blazars and extend our calculations of the 
extragalactic \gray\ background from blazars to an energy of 500 GeV with
absorption effects included (see section 4.3). 
Their results indicate that the extragalactic
\gray\ background spectrum from blazars should steepen significantly 
above 20 GeV, owing to extragalactic absorption. Future observations of
such a steepening would thus provide a test of the blazar origin
hypothesis for the \gray\ background radiation. The results of the SS98
absorption calculations can be used to place limits on the redshifts
of \gray\ bursts (see section 4.4). We describe and discuss these results
in the following subsections.
      
\subsection{Redshift Dependence of the Intergalactic Low Energy SED}

The opacity of intergalactic space to high energy
\grays\ as a function of redshift depends upon the 
number density of soft target photons 
(IR to UV) as a function of redshift,
photons whose production is dominated by stellar emission.  To evaluate the
SED of the IR-UV intergalactic radiation field
we must integrate the total stellar emissivity 
over time.  This requires an
estimate of the dependence of stellar emissivity on redshift. Previous 
calculations of \gray\ opacity
have either assumed that essentially all of the background was in 
place at high redshifts, corresponding to a burst of star 
formation at the initial redshift \cite{ste96}, \cite{sd96}, \cite{mp96}
or strong evolution \cite{mph96}, or that there is no evolution
\cite{mph96}. 

Pei and Fall \cite{pf95} have devised a method for calculating stellar 
emissivity which bypasses the uncertainties associated with estimates of 
poorly defined luminosity distributions of evolving galaxies.
The core idea of their 
approach is to relate the star formation rate 
directly to the evolution of the neutral gas density in damped
\lya\ systems, and then to use stellar population synthesis models to
estimate the mean co-moving stellar emissivity ${\cal E}_{\nu}(z)$
of the universe as a function of frequency $\nu$ and
redshift $z$ \cite{fall96}.
The SS98 calculation of stellar emissivity closely follows this 
elegant analysis, with minor modifications.

Damped \lya\ systems are high-redshift clouds of gas whose neutral
hydrogen surface density is large enough ($>2\times 10^{20}$ cm$^{-2}$)
to generate saturated \lya\
absorption lines 
in the spectra of background quasars that 
happen to lie along and
behind  common lines of sight to these clouds.  
These gas systems are believed
to be either precursors to galaxies or young galaxies themselves, 
since their neutral hydrogen (HI)
surface densities are comparable to those of spiral galaxies today, and their
co-moving number densities are consistent with those of present-day galaxies
\cite{wo86}, \cite{pe94}.  It is in these systems that initial
star formation presumably took place, so there is a relationship between
the mass content of stars and of gas in these clouds; if there is no infall
or outflow of gas in these systems, the systems are ``closed'', so that the
formation of stars must be accompanied by a reduction in the neutral gas
content.  Such a variation in the HI surface densities of
\lya\ systems with redshift is seen, and is used by Pei and Fall \cite{pf95}
to estimate the mean cosmological rate of star formation back to 
redshifts as large as $z=5$.

Pei and Fall \cite{pf95} estimated the
neutral (HI plus HeI) co-moving gas density 
$\rho_{c}\Omega_{g}(z)$ in damped \lya\ systems 
from observations of the redshift evolution of 
these systems by Lanzetta, \etal\ \cite{lan95}      . 
Lanzetta, \etal\ have observed 
that while the number density of damped \lya systems 
appears to be relatively constant over redshift, the fraction of higher
density absorption systems within this class of objects
decreases steadily with decreasing redshift.  They attribute this to a
reduction in gas density with time, roughly of the form
$\Omega_{g}(z)=\Omega_{g0}e^{z}$, where 
$\rho_{c}\Omega_{g0}$ is the 
current gas density in galaxies.  Pei and Fall have
taken account of self-biasing effects to obtain a corrected
value of $\Omega_{g}(z)$. SS98 \cite{ss98} have reproduced
their calculations to obtain $\Omega_{g}(z)$ under the assumptions that
the asymptotic, high redshift value of 
the neutral gas mass density is $\Omega_{g,i}=
1.6\times10^{-2}h_{0}^{-1}$, where $h_{0}\equiv 
H_{0}$/(100 km s$^{-1}$Mpc$^{-1}$).
In a ``closed galaxy'' model, the change in co-moving stellar mass density 
$\rho_{c}\dot{\Omega}_{s}(z)=-\rho_{c}\dot{\Omega}_{g}(z)$, since the gas 
mass density $\rho_{c}\Omega_{g}(z)$ is
being converted into stars. This determines the star formation
rate and consequent stellar emissivity.
The rate of metal production, $\dot{Z}$, is related to star formation
rate by $\Omega_{g}\dot{Z}=\zeta\dot{\Omega}_{s}$, 
where $\zeta=0.38 Z_{\odot}$ 
is the metallicity yield averaged over the initial stellar mass function, with
$Z_{\odot}$ being the solar metallicity \cite{pf95}.  This
gives a metallicity evolution $Z(z)=-\zeta\ln [\Omega_{g}(z)/\Omega_{g,i}]$.

In order to determine the mean stellar emissivity from the star
formation rate, an initial mass function (IMF) $\phi(M)$
must be assumed for the distribution of stellar masses $M$ in a
freshly synthesized stellar population. To further specify the luminosities
of these stars as a function of mass $M$ and age $T$, 
Fall, Charlot, and Pei \cite{fall96} use the Bruzual-Charlot (BC)
population synthesis models
for the spectral evolution of stellar populations \cite{br93}, \cite{ch91}.
In these population synthesis models, the
specific luminosity
$L_{\rm star}(\nu,M,T)$, of a star of mass $M$ and age
$T$ is integrated over a specified IMF to obtain a total specific
luminosity $S_{\nu}(T)$ per unit mass 
for an entire population,
in which all stellar members are produced 
simultaneously ($T=0$).  Following Fall, Charlot, and Pei \cite{fall96},
SS98 used the BC model corresponding to a Salpeter IMF,
$\phi(M)\,dM\propto M^{-2.35}\,dM$, where $0.1M_{\odot}<M<125M_{\odot}$.
The mean co-moving emissivity ${\cal E}_{\nu}(t)$
was then obtained by convolving over time $t$ 
the specific luminosity 
with the mean co-moving mass rate of star formation.
SS98 also obtained metallicity correction
factors for stellar radiation at various wavelengths. Increased metallicity 
gives a redder population spectrum \cite{wo94}, \cite{be94}.

SS98 calculated stellar
emissivity as a function of redshift at 0.28 $\mu$m, 0.44 $\mu$m,
and 1.00 $\mu$m, both with and without a metallicity correction.
Their results agree well with the
emissivity obtained by the Canada-French
Redshift Survey \cite{li96} over the redshift range of the observations
($z \le 1$).

The stellar emissivity in the universe is found to peak
at $ 1 \le z \le 2$, dropping off steeply at lower reshifts and more slowly
at higher redshifts. Indeed,
Madau, \etal\ \cite{mad96} have used observational data from the 
Hubble Deep Field to show that metal production has a similar redshift 
distribution, such production being a direct measure of the star formation 
rate. (See also Ref. \cite{st98}).

The co-moving radiation energy density $u_{\nu}(z)$ 
is the time integral of the co-moving emissivity ${\cal E}_{\nu}(z)$,
\begin{equation} \label{B}
u_{\nu}(z)=
\int_{z}^{z_{\rm max}}dz^{\prime}\,{\cal E}_{\nu^{\prime}}(z^{\prime})
\frac{dt}{dz}(z^{\prime})e^{-\tau_{\rm eff}(\nu,z,z^{\prime})},
\end{equation}
where $\nu^{\prime}=\nu(1+z^{\prime})/(1+z)$ and $z_{\rm max}$ is the
redshift corresponding to initial galaxy formation.
The extinction term $e^{-\tau_{\rm eff}}$ 
accounts for the absorption of ionizing photons by the clumpy
intergalactic medium (IGM) that lies between the source and observer.
Although the IGM is effectively transparent to non-ionizing photons,
the absorption of photons by HI,
HeI and HeII can be considerable \cite{mad95}.

\subsection{The Gamma-Ray Opacity at High Redshifts}

With the co-moving energy density $u_{\nu}(z)$ evaluated \cite{ss98} (SS98), 
the optical depth for \grays\ owing to electron-positron pair production 
interactions with photons of the stellar radiation
background can be determined from the expression \cite{sds92}

\begin{equation} \label{G}
\tau(E_{0},z_{e})=c\int_{0}^{z_{e}}dz\,\frac{dt}{dz}\int_{0}^{2}
dx\,\frac{x}{2}\int_{0}^{\infty}d\nu\,(1+z)^{3}\left[\frac{u_{\nu}(z)}
{h\nu}\right]\sigma_{\gamma\gamma}(s)
\end{equation}
where $s=2E_{0}h\nu x(1+z)$,
$E_{0}$ is the observed \gray\ energy at redshift zero, 
$\nu$ is the frequency at redshift $z$,
$z_{e}$ is the redshift of
the \gray\ source, $x=(1-\cos\theta)$, $\theta$ being the angle between
the \gray and the soft background photon, $h$ is Planck's constant, and
the pair production cross section $\sigma_{\gamma\gamma}$ is zero for
center-of-mass energy $\sqrt{s} < 2m_{e}c^{2}$, $m_{e}$ being the electron
mass.  Above this threshold, 
\begin{equation} \label{H}
\sigma_{\gamma\gamma}(s)=\frac{3}{16}\sigma_{\rm T}(1-\beta^{2})
\left[ 2\beta(\beta^{2}-2)+(3-\beta^{4})\ln\left(\frac{1+\beta}{1-\beta}
\right)\right],
\end{equation}
where $\beta=(1-4m_{e}^{2}c^{4}/s)^{1/2}$.

\begin{figure}[t]
\centerline{\psfig{file=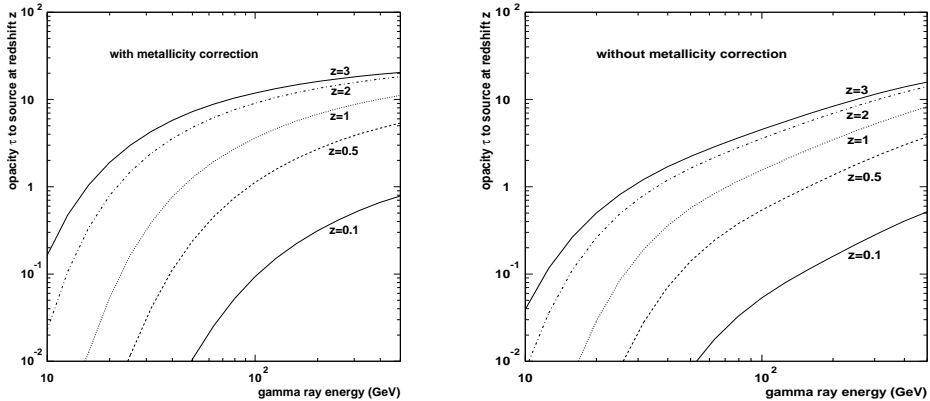,width=13.0truecm}}
\caption{ The opacity $\tau$ of the universal soft photon background to
\grays\ as a function of \gray\ energy and source redshift (from SS98)
\protect\cite{ss98}.  
These curves are calculated with and without a metallicity correction.}
\label{opac}

\end{figure}

Figure \ref{opac}
shows the opacity $\tau(E_{0},z)$ for the energy
range 10 to 500 GeV, calculated by SD98 both with and without a 
metallicity correction.
Extinction of \grays\ is negligible below 10 GeV.
~

The weak redshift dependence of the opacity at the higher redshifts 
as shown in Figure ~\ref{opac}  indicates that the opacity is not very
sensitive to the initial epoch of galaxy formation, contrary to the 
speculation of MacMinn and Primack \cite{mp96}. In fact, the uncertainty in the
metallicity correction (see Figure \ref{opac}) would obscure any dependence on
$z_{max}$ even further.

\subsection{The Effect of Absorption on the Spectra of Blazars and the
Gamma-Ray Background}

\begin{figure}[t]
\centerline{\psfig{file=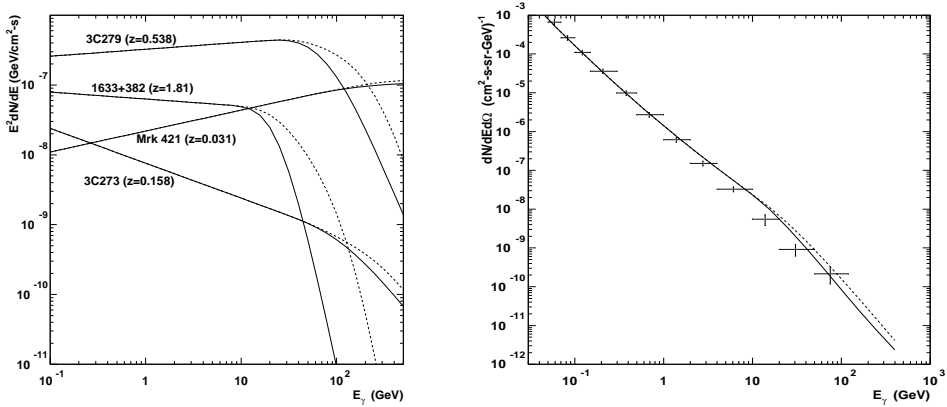,width=13.0truecm}}
\caption{ The left graph shows the effect of intergalactic absorption by 
pair-production on the power-law spectra of
four prominent grazars: 1633+382 ($z=1.81$), 
3C279 ($z=0.54$), 3C273 ($z=0.15$), and Mrk 421 ($z=0.031$); The right
graph shows the
extragalactic \gray\ background spectrum predicted by the unresolved 
blazar model of
Ref. \protect\cite{ss96} with absorption included, calculated for a mean 
{\it EGRET} point-source 
sensitivity of $10^{-7}$ cm$^{-2}$s$^{-1}$, compared with the {\it EGRET}
data on the \gray\ background \protect\cite{sr98}.  
The solid (dashed) curves are calculated with (without) the metallicity
correction function (from SS98 \protect\cite{ss98}).}
\label{blaz}
\end{figure}

With the \gray\ opacity $\tau(E_{0},z)$ calculated out to
$z=3$,
the cutoffs in blazar \gray\ spectra caused by extragalactic pair 
production interactions with stellar photons can be predicted.
The left graph in Figure \ref{blaz} from Ref. \cite{ss98} (SS98)
shows the effect of the intergalactic radiation
background on a few of the grazars 
observed by {\it EGRET},
{\it viz.}, 1633+382, 3C279, 3C273, and Mrk 421,
assuming that the mean spectral indices obtained for these sources by
{\it EGRET} extrapolate out to higher energies 
attenuated only by intergalactic 
absorption.  Observed cutoffs in grazar spectra 
may be intrinsic cutoffs in \gray production in
the source, or may be caused
by intrinsic \gray\ absorption within the source itself. 

~

The right hand graph in Figure \ref{blaz}
shows the background spectrum predicted from unresolved blazars
\cite{ss96}, \cite{ss98}
compared with the {\it EGRET} data \cite{sr98}. Note that the 
predicted spectrum steepens 
above 20 GeV, owing to extragalactic absorption by pair-production 
interactions with radiation from external galaxies, particularly at
high redshifts.
Above 10 GeV, blazars may have natural cutoffs in
their source spectra \cite{sds96} and intrinsic 
absorption may also be important in some sources \cite{pb96}. 
Thus, above 10 GeV the 
calculated background flux from unresolved blazars shown in 
Figure \ref{blaz} may actually be an upper limit.
Whether cutoffs in grazar spectra are 
primarily caused by intergalactic absorption can be determined by
observing whether the grazar cutoff energies have the 
type of redshift dependence predicted here.

\subsection{Constraints on Gamma-ray Bursts}

The discovery of optical and X-ray afterglows of \gray\ bursts and 
the identification of host galaxies with measured redshifts, \ie\ ,
\cite{me97}, \cite{ku98}, has lead the accumulation of evidence that
these bursts are highly relativistic fireballs originating at cosmological
distances \cite{liv98} and may be associated primarily with early star
formation \cite{dj98}.

As indicated in Figure \ref{opac}
\grays\ above an energy of $\sim$ 15 GeV will
be attenuated if they at emitted at a redshift of $\sim$ 3.
On 17 February 1994, the {\it EGRET} telescope observed a \gray\ burst
which contained a photon of energy $\sim$ 20 GeV \cite{hu94}.
As an example, if one adopts the opacity results which include the
metallicity correction, the highest energy photon in this burst 
would be constrained probably
to have originated at a redshift
less than $\sim$2. 
Future detectors such as {\it GLAST} (Ref. \cite{bl96}, also Gehrels, these 
proceedings) may be able to place better redshift 
constraints on bursts observed at higher energies. 
Such constraints may further help to identify the host galaxies of \gray\
bursts.

\section{Acknowledgment}

I wish to acknowledge that the work presented here was a result of extensive 
collaboration with O.C. De Jager, M.A. Malkan, and M.H. Salamon, as indicated 
in the references cited. I also wish to thank Okkie De Jager for helping with 
the manuscript.

\end{document}